\documentclass[10pt]{article}   
\usepackage{opex3}
\usepackage{graphicx}
\usepackage{bm}
\graphicspath{{pict/}{}}
\usepackage{amsmath,amssymb}

\begin{document}

\title{First-principles method for high-$Q$ photonic crystal cavity mode calculations}

\author{Sahand Mahmoodian$^{*,1}$, J.E. Sipe$^2$, Christopher G. Poulton$^{3}$, Kokou B. Dossou$^{3}$, Lindsay C. Botten$^{3}$, Ross C. McPhedran$^{1}$, and C. Martijn de Sterke$^{1}$}
\address{
$^1$CUDOS and IPOS, School of Physics, University of Sydney, Australia\\
$^2$Dep. of Physics, University of Toronto, 60 St. George Street, Toronto, ON M5S 1A7, Canada\\
$^3$CUDOS, School of Mathematical Sciences, University of Technology, Sydney, Australia\\
$^*$Corresponding author: sahand@physics.usyd.edu.au}

\begin{abstract}
We present a first-principles method to compute radiation properties of ultra-high quality factor photonic crystal cavities. Our Frequency-domain Approach for Radiation (FAR) can compute the far-field radiation pattern and quality factor of cavity modes $\sim\!100$ times more rapidly than conventional finite-difference time domain calculations. It also provides a simple rule for engineering the cavity's far-field radiation pattern.
\end{abstract}


The high quality factors ($Q$) and small modal volumes of photonic crystal (PC) cavities make them ideally suited for applications requiring strong optical field enhancement, such as low-energy optical switching \cite{nozaki2010sub}, strongly coupled cavity quantum electrodynamics (QED) \cite{yoshie2004vacuum} and harmonic generation \cite{mccutcheon2007experimental}. Recent interest has focused on high-$Q$ cavities where the far-field radiation pattern is engineered to emit vertically, enabling free-space mode excitation \cite{PhysRevB.82.075120,englund2010resonant,portalupi2010planar} in cavity QED \cite{englund2010resonant} and harmonic generation \cite{galli2010low} experiments.

Photonic crystal cavity design uses established theoretical ideas \cite{johnson2001multipole,akahane2003high, song2005ultra,sauvan2005modal,englund2005general} to maximize quality factors, in conjunction with finite difference time domain (FDTD) calculations to compute the cavity mode. Even with improvements in speed and accuracy \cite{mandelshtam1997harmonic,farjadpour2006improving}, time domain calculations are by nature computationally intensive for the long life-times of ultra-high $Q$ cavities, taking hours or days per design on a supercomputer. Optimizing both the quality factor and the radiation pattern can require the exploration of a large parameter space \cite{portalupi2010planar,englund2010resonant}, further increasing the computation effort. These severe computational demands, and the inability of FDTD to provide insight into the underlying physics, point to the need for an alternative method. Here we provide such a method. Our first-principles Frequency-domain Approach for Radiation (FAR) is $\sim\!100$ times more efficient than FDTD calculations since we do not compute radiative modes directly. It consists of two parts: we initially approximate the cavity mode as a bound mode after which the radiation is obtained using perturbation theory, thus avoiding the most time-consuming part of FDTD calculations. The FAR also provides a design strategy for achieving cavities with specified radiation patterns without requiring exhaustive simulations.

\begin{figure}[!t]
\centering\includegraphics[width = 0.8\textwidth]{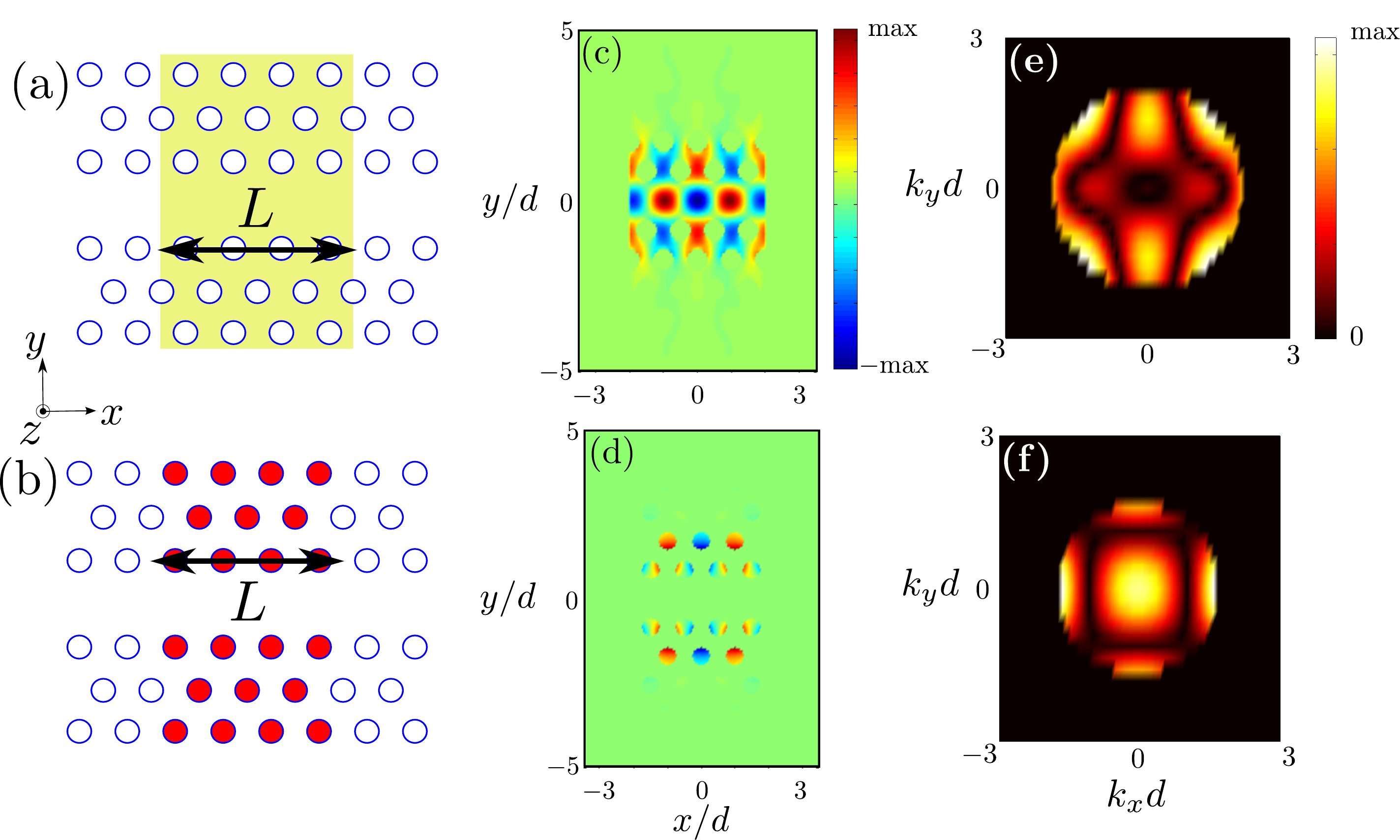}
\caption{\label{GammaTildeDy} Schematic of (a) photosensitive cavity with locally increased background index and of (b) fluid infiltrated cavity with increased hole index.  $\tilde{A}(\mathbf r)\mathbf D^a(\mathbf r)$ (arb. units) for (c) photosensitive cavity (d)  fluid infiltrated cavity, both with a length of $L=4d$. (e)-(f) Modulus of the Fourier transform of (c)-(d) respectively, with non-radiating components removed.}
\end{figure}

We apply the FAR to double heterostructure photonic crystal cavities \cite{song2005ultra}, which are formed by perturbing a photonic crystal waveguide (PCW) in a slab geometry. The two geometries are shown in Fig.~\ref{GammaTildeDy}(a)-(b); in the {\it photosensitive cavity} (Fig. \ref{GammaTildeDy}(a)), the refractive index of a strip around the PCW (yellow shading) is uniformly increased by $\Delta n_p$, as can be achieved in chalcogenide glass \cite{tomljenovic2007high,lee2009photowritten}.  In the {\it fluid infiltrated cavity} (Fig.~\ref{GammaTildeDy}(b)), the refractive index of the holes is increased by $\Delta n_i$ in a strip-like region (red shading), typically by fluid infiltration \cite{tomljenovic2006design,bog2008high}.  These two cavities are therefore complementary, {\sl i.e.}, in one only the background is perturbed, while in the other only the holes. We have found that  considerable qualitative insight into the radiation pattern of the cavity mode can be obtained by examining a single term in the equation that governs the radiation from the cavity. This term has the form  $\tilde{A}(\mathbf r)\mathbf D^a(\mathbf r)$, where $\tilde{A}(\mathbf r)$ is associated with the perturbation that creates the cavity and $\mathbf D^a(\mathbf r)$ is the bound approximation for the cavity mode. This term is shown for a $z=0$ slice through the PC slab in Figs. \ref{GammaTildeDy}(c)-(d). The perturbation term $\tilde{A}(\mathbf r)$ is only non-zero in the background for the photosensitive cavity (Fig. \ref{GammaTildeDy}(c)) and only non-zero in the holes for the fluid infiltrated cavity (Fig. \ref{GammaTildeDy}(d)). The Fourier components in the light cone of this product are peaked near the edge of the light cone for the photosensitive cavity (Fig. \ref{GammaTildeDy}(e)) corresponding to radiation being directed towards the horizon. However, for the fluid infiltrated cavity, the Fourier transform is strongest near $k_x=k_y=0$ (Fig. \ref{GammaTildeDy}(f)), and thus it predominantly radiates vertically. We later return to this insight and use it to provide a general design rule for engineering the radiation pattern of a cavity mode.

Our theory uses a Hamiltonian formulation \cite{chak2007hamiltonian} to construct cavity modes by superposing a basis of bound PCW modes expressed in terms of the $\mathbf B(\mathbf r)$ and  $\mathbf D(\mathbf r)$ fields, so any superposition is divergence-free. The Hamiltonian for a dielectric PC cavity with relative permittivity $\epsilon(\mathbf r)$ is
   \begin{equation}
   \label{hamiltonian}
   {\cal H} = \frac{1}{2\mu_0}\int\! d\mathbf r\, \mathbf B(\mathbf r) \cdot \mathbf B(\mathbf r) + \frac{1}{2\epsilon_0}\int\! d\mathbf r\, \frac{\mathbf D(\mathbf r)\cdot \mathbf D(\mathbf r)}{\epsilon(\mathbf r)}.
   \end{equation}
Since we use PCW modes as a basis, it is convenient to define $\epsilon(\mathbf r) = \bar{\epsilon}(\mathbf r) + \tilde{\epsilon}(\mathbf r)$, where $\bar{\epsilon}(\mathbf r)$ is the permittivity of the PCW, and $\tilde{\epsilon}(\mathbf r)$ the small permittivity change that creates the cavity. We then expand the cavity mode using the normalized PCW modes \cite{chak2007hamiltonian} below the light cone
   \begin{equation}
   \label{expansion}
   \mathbf D(\mathbf r,t) = \int_{\rm bound}\!\! d k\, \sqrt{\frac{\hbar \omega_k}{2}}\,a_k\, e^{-i\omega_k t} \mathbf D_k(\mathbf r) + c.c.,
   \end{equation}
where we only include modes of the even PCW band, i.e. those for which $E_y(\mathbf r)$ is even in $y$. We can include more modes, but ultra-high $Q$ cavity modes are typically gently confined and thus different bands couple weakly.

Substituting (\ref{expansion}) into (\ref{hamiltonian}) we obtain an approximation for the Hamiltonian of the PC cavity
   \begin{equation}
    \label{pertHamilt}
   {\cal H}_{1} = \int dk dk' \left[ \hbar \omega_k \delta(k -k') +\hbar\sqrt{\omega_k\omega_{k'}}\int d \mathbf r \, \gamma(\mathbf r)\,\mathbf D_k^*(\mathbf r)\cdot \mathbf D_{k'}(\mathbf r) \right] a^\dag_k a_{k'}
   \end{equation}
where $\gamma(\mathbf r) = 1/(2\epsilon_0)\left[1/\epsilon(\mathbf r) - 1/\bar{\epsilon}(\mathbf r)\right]$, and we dropped non-rotating wave terms involving $a^\dag_k a^\dag_{k'}$ and $a_k a_{k'}$. Diagonalizing ${\cal H}_1$ determines an eigenvalue, the energy $\hbar \omega_0$ of a photon in the cavity mode, while its eigenfunction $v_0(k)$ gives the cavity mode in the basis of PCW modes:
   \begin{equation}
   \label{Dapprox}
   \mathbf D^a(\mathbf r) = \int_{\rm bound} dk \sqrt{\frac{\hbar \omega_k}{2}}v_0(k)\mathbf D_k(\mathbf r).
   \end{equation}
We now have an approximate expression for the cavity mode in terms of a basis with Fourier components outside the light cone. The Fourier content within the light cone of the ultra-high $Q$ factor cavities of interest here is small, and we have found that $\mathbf D^a(\mathbf r)$ is a good approximation for the shape of the cavity mode. Similarly, the eigenvalues of (\ref{pertHamilt}) approximate the real part of the frequency of the cavity mode well. We thus use $\mathbf D^a(\mathbf r)$ to find a first approximation for the polarization field $\mathbf P(\mathbf r)$ within the light cone.


The polarization field $\mathbf P(\mathbf r)= \epsilon_0 \left[ \epsilon(\mathbf r) - 1\right] \mathbf E(\mathbf r)$ of a mode with frequency $\omega$ satisfying the macroscopic Maxwell equations is also a solution to the integral equation
\begin{equation}
\label{greens}
\mathbf P(\mathbf r) =  \epsilon_0 \left[\epsilon(\mathbf r) - 1 \right]\int d \mathbf r' G(\mathbf r - \mathbf r'; \omega)\mathbf P (\mathbf r'),
\end{equation}
where the Green tensor expresses the electric field at $\mathbf r'$ due to an oscillating polarization source at $\mathbf r$. We use the formalism for layered media \cite{sipe1987new}, in which we deal with a sheet of polarization. Since we need to compute the out-of-plane ($z$-direction) radiation of a PC cavity, this formalism is particularly useful as it separates propagating modes in the $z$-direction, with $|\bm \kappa|^2 \equiv k_x^2 + k_y^2 \leq k_0^2$, from evanescent modes with $|\bm{\kappa}|^2 > k_0^2$, where $k_0=\omega_0/c$.

Defining $\Gamma(\mathbf r) = \left[\epsilon(\mathbf r) - 1\right]/\epsilon(\mathbf r)$, the polarisation field of the cavity is approximated by $ \mathbf P^a(\mathbf r) = \Gamma(\mathbf r)\mathbf D^a(\mathbf r) \equiv (\bar{\Gamma}(\mathbf r) + \tilde{\Gamma}(\mathbf r) )\mathbf D^a(\mathbf r)$, where again the over-bar denotes a quantity for the PCW and the tilde denotes the perturbation creating the cavity. Since $\mathbf D^a(\mathbf r)$ has no Fourier components within the light cone, neither does $\bar{\Gamma}(\mathbf r)\mathbf D^a(\mathbf r)$, $\bar{\Gamma}(\mathbf r)$ being periodic with the lattice. However, $\tilde{\Gamma}(\mathbf r)\mathbf D^a(\mathbf r)$ does have components within the light cone, providing a starting point for calculating the radiative polarization. We relate the actual polarization field of the cavity mode $\mathbf P(\mathbf r)$ to $\mathbf P^a(\mathbf r)$ by writing $\mathbf P(\mathbf r) = \mathbf P^a(\mathbf r) + \mathbf P_c(\mathbf r)$, whose radiative components are $\mathbf P^{\rm rad}(\mathbf r) = \tilde{\Gamma}(\mathbf r)\mathbf D^a(\mathbf r)+ \mathbf P_c^{\rm rad}(\mathbf r)$, where $\mathbf P_c(\mathbf r)$ is the correction to the polarization field, while {\sl rad} refers only to Fourier components in the light cone. We write the complex cavity mode frequency $\omega$ as $\omega = \omega_0 + \tilde{\omega}$. We next perform a Taylor expansion about $\omega_0$ of the Green function, substitute into (\ref{greens}), and use the fact that $\mathbf P_c^{\rm rad}(\mathbf r)$ and the variables with tildes are small. After some manipulation and keeping only terms with Fourier components in the light cone, we obtain a first order expression for $\mathbf P^{\rm rad}$
\begin{equation}
\label{master}
\mathbf P^{\rm rad}_1(\mathbf r) - \epsilon_0\left(\bar{\epsilon}(\mathbf r) - 1 \right)\!\!\int\!\! d\mathbf r' G(\mathbf r - \mathbf r';\omega_0)\mathbf P^{\rm rad}_1(\mathbf r')=\tilde A(\mathbf r)\mathbf D^a(\mathbf r)\equiv\left[\tilde{\Gamma}(\mathbf r) + \frac{\tilde{\epsilon}(\mathbf r)\bar{\Gamma}(\mathbf r)}{\bar{\epsilon}(\mathbf r)} \right]\!\mathbf D^a(\mathbf r),
   \end{equation}
where the driving term, which contains information about the cavity via the parameters with a tilde, couples to Fourier components inside the light cone. As discussed earlier in this paper, we have found that $\tilde A(\mathbf r)\mathbf D^a(\mathbf r)$ gives good qualitative insight into the far-field radiation. In general though, Eq.~(\ref{master}) is a Fredholm integral equation of the second kind, in which the Green function ensures a self-consistent interaction between the dipoles.

By solving (\ref{master}), we obtain the full quantitative radiative polarization components of the cavity mode, from which the far-field radiation can be determined using the Green function in~(\ref{greens}). In the far-field we write the electric field as $\mathbf E_{\rm far}(\mathbf r) = \mathbf e_{\pm}(\bar{\bm{\kappa}})e^{i k_0 r}/r$, where $\bar{\bm{\kappa}} \equiv k_0 {\hat{\mathbf r}}\cdot(\hat{\mathbf x}\hat{\mathbf x} + \hat{\mathbf y}\hat{\mathbf y})$, with, above $(+)$ and below $(-)$ the slab,
\begin{equation}
\label{slabSum}
e_\pm^s(\bm{\kappa}) = \frac{k_0^2}{4 \pi \epsilon_0}\,\hat{\mathbf s}\cdot \int dz\, d \mathbf R\, e^{-i\bm{\kappa}\cdot \mathbf R}e^{\mp iwz}\,\mathbf P^{\rm rad}_1(\mathbf R,z)\\
   \end{equation}
for $s$ polarization, and with a similar expression for $p$ polarization. Here $\mathbf R = (x,y)$ and $w = (k_0^2 - |\bm{\kappa}|^2)^{1/2}$. Equation~(\ref{slabSum}) is thus a planar ($x$ and $y$) Fourier Transform, integrated over the thickness of the slab ($z$) with appropriate phases. Each $(k_x,k_y)$ of the polarization field inside the light cone corresponds to a unique far-field direction. The far-field electric field gives the Poynting vector, and therefore the quality factor of the cavity mode can be computed.

\begin{figure}[!t]
\centering\includegraphics[width=0.8\textwidth]{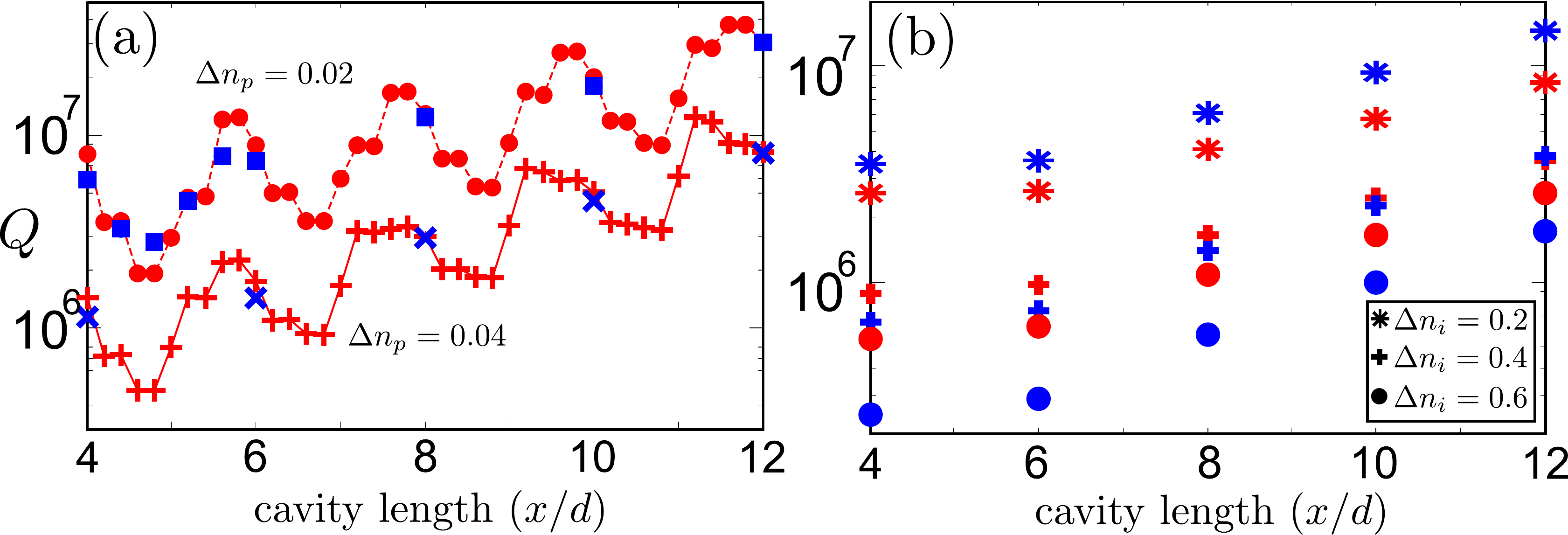}
\caption{\label{doubleQs} Quality factor versus cavity length for (a) the photosensitive cavity (Fig.~\ref{GammaTildeDy}(a)); (b), the fluid infiltrated cavity (Fig.~\ref{GammaTildeDy}(b)). Red symbols are computed using the FAR method, while blue ones are computed by FDTD.}
\end{figure}

To obtain numerical solutions to (\ref{master}), we further assume that Fourier components inside the light cone do not couple to those outside the light cone. Since inside the light cone $k_{x,y}$ are small, this lets us use a coarse discretization in $x$ and $y$, reducing the size of the problem. By using an efficient iterative bi-conjugate gradient method \cite{zhang1997gpbi,chaumet2009efficient}, Eq.~(\ref{master}) can be solved to within a tolerance of $10^{-5}$ in $20-100$ iterations, each of which take less than 10~seconds. Our MATLAB code typically solves (\ref{pertHamilt}) and (\ref{master}) in under 15 minutes on a work station. In contrast, the FDTD calculations for each point in Fig.~\ref{doubleQs} took tens of hours on a 32 core cluster.

In our simulations for the photosensitive cavity (Fig.~\ref{GammaTildeDy}(a)), we take a $W1$ PCW with background index of $n_b=2.7$, slab thickness, $t=0.7d$ and hole radius $a=0.3d$, where $d$ is the period and $\Delta n_p=0.02, 0.04$. For the fluid infiltrated cavity (Fig.~\ref{GammaTildeDy}(b)) we use a $W0.98$ silicon PCW (background index $n_b=3.46$), with slab thickness, $t=0.49d$, hole radius $a=0.26d$, $\Delta n_i=0.2, 0.4, 0.6$. In Fig.~\ref{doubleQs} we show the $Q$-factor versus cavity length calculated using the FAR method (red) and using FDTD (blue). In Fig.~\ref{doubleQs}(a), which is for photosensitive cavities, the efficiency of our theory allows us to vary the cavity length continuously. This is impractical for FDTD calculations, so we only have results at even integer values of the cavity length and at some intervening points. The agreement between the results is excellent: the $Q$-factors agree to within 30\% (or their logarithms by 2\%), making them suitable for examining trends in $Q$. The strong oscillations in $Q$ correspond to a factor of $8$. In Fig.~\ref{doubleQs}(b), which is for fluid infiltrated cavities, we only calculated $Q$ for even integer cavity lengths. The agreement for these cavities is good: the results have the same trends and never differ by more than a factor two.

\begin{figure}[!th]
\centering\includegraphics[width=0.8\textwidth]{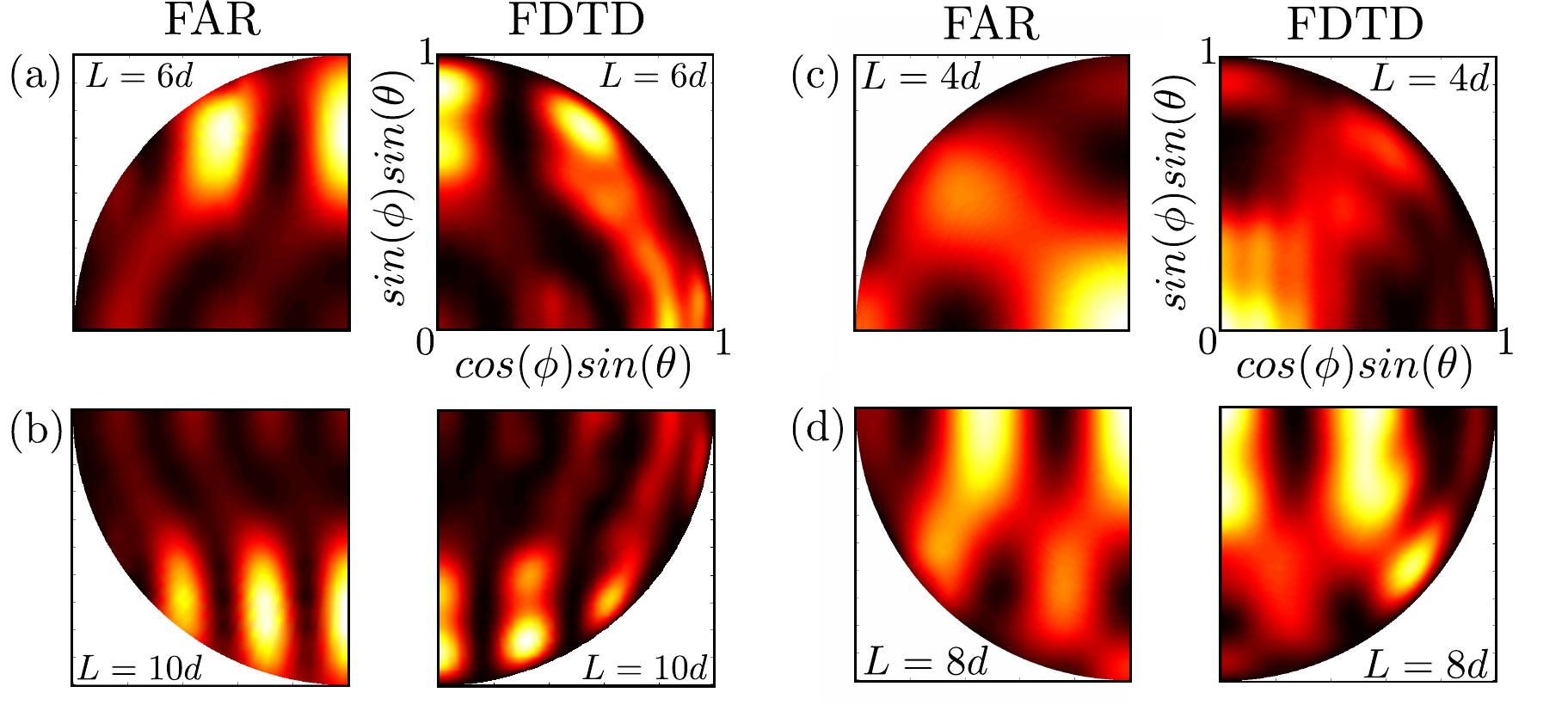}
  \caption{\label{farFields} Symmetric quadrants of far-field radiation ($S_r$) for (a),(b) cavities in Fig. \ref{GammaTildeDy}(a) with $\Delta n_p=0.02$ and (c),(d) those in Fig. \ref{GammaTildeDy}(b) with $\Delta n_i=0.2$. Left frames are computed using the FAR method while right frames are computed using FDTD. Colors are as in Fig. \ref{GammaTildeDy}(d). Angles $\phi$ and $\theta$ are azimuthal and declination angles respectively.}
\end{figure}

Having demonstrated the reliability of the FAR, we now exploit its semi-analytic nature to design desirable far-field radiation properties. Figure~\ref{farFields} shows good agreement between the far-field radiation patterns computed using the FAR (left) and FDTD (right), for photosensitive cavities (Figs.~\ref{farFields}(a),(b)) and fluid infiltrated cavities (Figs.~\ref{farFields}(c),(d)) of different lengths $L$. Note that (i) the number of lobes in the radiation pattern increases as the cavity gets longer; and that (ii) as discussed earlier, photosensitive cavities radiate predominantly at large declination angles ($\theta$), while the fluid infiltrated cavities radiate mostly vertically. Both features can be explained by examining $\tilde{A}(\mathbf r) \mathbf D^{a}(\mathbf r)$. The effect of  $\tilde{A}(\mathbf r)$ on $\mathbf D^{a}(\mathbf r)$ is to introduce nodes and anti-nodes due to Fabry-Perot effects in the cavity. Point (ii) is more subtle: returning to Fig. \ref{GammaTildeDy}, since the cavity modes are dielectric modes,  for the photosensitive cavity the product $\tilde{A}(\mathbf r) \mathbf D^{a}(\mathbf r)$ (Fig. \ref{GammaTildeDy}(c)) has the effect of merely introducing sidelobes in the Fourier transform of $\mathbf D^a(\mathbf r)$. The overlap of these sidelobes with the light cone (Fig. \ref{GammaTildeDy}(e)), is dominated by $(k_x,k_y)$ values at the edge of the light cone, maximizing the $Q$ factor and leading to radiation at large declination angles.

For fluid infiltrated cavities, $\tilde{A}(\mathbf r) \mathbf D^{a}(\mathbf r)$ (Fig.~\ref{GammaTildeDy}(d)), is nonzero only inside holes. Its Fourier transform within the light cone (Fig.~\ref{GammaTildeDy}(f)) peaks at the origin, because the cavity length is such that $\tilde{A}(\mathbf r) \mathbf D^{a}(\mathbf r)$ has a strong non-zero $DC$ Fourier component. This is clear from the fields in the holes in Fig.~\ref{GammaTildeDy}(d): four holes have strong positive fields and only two have strong negative fields because the cavity mode is dominated by the Bloch mode at $k d = \pi$, which changes sign each period. This cavity therefore radiates mostly vertically. Since examining $\tilde{A}(\mathbf r) \mathbf D^{a}(\mathbf r)$ is sufficient for qualitative insight into the far-field, the requirement for steering the radiation of cavity modes is thus simple: construct a perturbation such that $\tilde{A}(\mathbf r)\mathbf D^a(\mathbf r)$ has a Fourier transform which peaks at $(k_x, k_y)$ values corresponding to the desired direction.

Similar arguments can be used to explain the variations in $Q$ observed in Fig.~\ref{doubleQs}(a): the cavity mode is a superposition of Bloch functions centred about the Brillouin zone edge ($k d=\pi$). It is thus not surprising that the period of the oscillations in Fig.~\ref{doubleQs}(a) corresponds to the period of the central Bloch function. The details of the oscillations in $Q$ depend on the superposition of the Bloch modes in $\tilde{A}(\mathbf r) \mathbf D^{a}(\mathbf r)$ overlapping with the light cone.

We have presented a frequency-domain approach for radiation (FAR) that allows the efficient calculation of the radiative properties of ultra-high $Q$ PC cavities. Both $Q$-factors and the radiation patterns are in good agreement with fully numerical FDTD calculations. The orders-of-magnitude improvement in computation speed will enable the application of powerful optimization algorithms, potentially transforming PC cavity design. The FAR lets us directly predict the radiation pattern through its link to the cavity's refractive index. Although we applied the theory to cavities created by refractive index changes, extensions allow the treatment of other cavity types, created, for example, by shifting inclusions \cite{kuramochi2006ultrahigh} or stretching the lattice \cite{song2005ultra}.

The authors thank A. Rahmani, M.J. Steel and C. Husko for discussions. This work was produced with the assistance of the Australian Research Council (ARC) under the ARC Centres of Excellence program, and was supported by the Flagship Scheme of the National Computational Infrastructure of Australia, and by the Natural Sciences and Engineering Research Council of Canada (NSERC).

\end{document}